\documentclass[useAMS,usenatbib]{mn2e}

\usepackage{epsfig}
\usepackage{subfigure}

\usepackage{amsmath}
\usepackage{amssymb}
\usepackage{graphicx}
\usepackage{latexsym}
\usepackage{aas_macros}

\usepackage{url}

\newcommand{\scinot}[2]{\ensuremath{#1 \times 10^{#2}}}

\newcommand{\paren}[1]{\left ( #1 \right )}

\newcommand{\parenfrac}[2]{\paren{\frac{#1}{#2}}}

\newcommand{\differd}[1]{\textrm{d}^{#1}}
\newcommand{\differ}[1]{\differd{}#1}

\newcommand{\deriv}[2]{\frac{\differ{#1}}{\differ{#2}}}

\newcommand{\Mstar}[1]{\ensuremath{M_{*}^{#1}}}
\newcommand{\Mplanet}[1]{\ensuremath{M_{\text{p}}^{#1}}}

\newcommand{\aplanet}[1]{\ensuremath{a_{\text{p}}^{#1}}}
\newcommand{\eplanet}[1]{\ensuremath{e_{\text{p}}^{#1}}}

\newcommand{\SigmaGap}[1]{\ensuremath{\Sigma_{\text{gap}}^{#1}}}

\begin{document}

\title[The formation of an eccentric gap]{The formation of an eccentric gap in a gas disk by a planet in an eccentric orbit}
\author[Hosseinbor et al.]{A. Pasha Hosseinbor,\thanks{aph5k@virginia.edu} Richard Edgar,\thanks{rge21@pas.rochester.edu} Alice C. Quillen,\thanks{aquillen@pas.rochester.edu}  \&  Amanda LaPage  \\
Department of Physics and Astronomy, University of Rochester, Rochester, NY 14627} 
\date{\today}

\pagerange {\pageref{firstpage}--\pageref{lastpage}}

\label{firstpage}


\maketitle

\begin{abstract}
We investigate the effect of a planet on an eccentric orbit on a two dimensional low mass gaseous disk.  
At a planet eccentricity above the planet's Hill radius divided by its semi-major axis,
we find that the disk morphology differs from that exhibited by a disk containing  a planet in a circular orbit.  
An eccentric gap is created with eccentricity that can exceed the planet's eccentricity and precesses with respect to the planet's orbit.   
We find that a more massive planet is required to open a gap when the planet is on an eccentric orbit.
We attribute this behavior to spiral density waves excited at corotation resonances that increase the disk's eccentricity and exert a torque opposite in sign to that exerted by the Lindblad resonances. 
The reduced torque makes it more difficult for waves driven by the planet to overcome viscous inflow in the disk.
Spectral energy distributions of transitional disks may reveal the presence of an eccentric planet if they are matched by an inner edge comprising a range of temperatures set by the semi-major axis and eccentricity of the hole.
\end{abstract}





\section{Introduction}
\label{sec:intro}

The recently discovered extra solar planets include Jovian mass planets with short period orbits and isolated planets with large eccentricities\footnote{\url{http://exoplanet.eu/}}.
Explanations for the ``hot Jupiters'' include orbital migration models \citep[e.g.][]{papa06}.
Explanations for the later include eccentricity growth via planet/disk interactions \citep[e.g.][]{2003ApJ...585.1024G,2004ApJ...606L..77S,2006A&A...447..369K}, or planet/planet interactions \citep[e.g.][]{2001MNRAS.320L..55M,2003ApJ...597..566T,2003CeMDA..87...53P,2004A&A...414..735K}.
Protoplanets could also be formed with moderate eccentricity due to interactions and waves in the disk \citep[e.g.][]{2006A&A...450..833C}.
Because protoplanets embedded in disks might be in eccentric orbits we are prompted to study gap formation by an eccentric planet.

So called ``transitional disks'' (systems with an accretion disc signature, where the accretion disc does not appear to extend to the star) have been observed for a number of years \citep{1992ApJ...395L.115M,1997AJ....114..301J,2003MNRAS.342...79R,2004ApJ...614L.133B}.
With the launch of the Spitzer Space Telescope, observers have been able to obtain high resolution spectra of these systems (previously, only broadband colours were available), dramatically improving our understanding of these systems.
In particular, Spitzer has established that the inner disks are very empty, and that the edges are quite sharp \citep{2005ApJ...621..461D,2005ApJ...630L.185C}.
These objects, such as CoKuTau/4, likely harbor massive planets residing just interior to the disk edge.
Interactions between the disk and planet and multiple planet interactions could cause this outer planet to be on an eccentric orbit.

In this paper using 2D hydrodynamic simulations, we investigate the ability of a planet on an eccentric orbit to open a gap in a low mass viscous disk.
We also investigate possible differences in disk morphology that are peculiar to planets on eccentric orbits and so  would allow an observer to differentiate between a disk perturbed by an eccentric planet and one in a circular orbit.

In section~\ref{sec:numeric} our simulations are described.
In section~\ref{sec:gapcriterion} we discuss how we identify a gap from the numerical simulations.
We discuss comparison numerical simulations of gaps opened by planets on circular orbits. 
In section~\ref{sec:eccgapdescribe} we discuss how a disk responds to a planet on an eccentric orbit.
In section~\ref{sec:eccgapcriterion} we present a modified gap opening criterion, based on our simulations, for planets on eccentric orbits. A summary and discussion follows.


\section{Numerical Experiments}
\label{sec:numeric}

To investigate the evolution of gaps opened by a single protoplanet, we carried out a set of two dimensional hydrodynamical calculations using the \textsc{Fargo} code developed by \citet{2000ASPC..219...75M,2000A&AS..141..165M}.
\textsc{Fargo} is an Eulerian polar grid code with a staggered mesh and an artificial second order viscous pressure to stabilize the shocks.
The code allows tidal interaction between one or more planets and a 2D non-self-gravitating gaseous disk
It owes its name to the \textsc{Fargo} advection algorithm that  removes the
average azimuthal velocity for the Courant timestep limit \citep{2000ASPC..219...75M}.
The simulations are performed in the frame rotating with the planet's guiding centre.
The outer boundary does not allow either inflow or outflow, so it must be located sufficiently far from the planet to ensure that spiral density waves are damped before they reach it.
This is facilitated by adopting a logarithmic grid in radius.
The grid inner boundary only allows material to escape so that the disk material may be accreted on to the primary star.
While the gas disk feels a gravitational perturbation from the planet, the planet itself was not allowed to feel the disk.  
This choice allows us to investigate the role of the planet's eccentricity in perturbing the disk without the additional complication of a varying planet eccentricity or inward migration.

The code uses units such that $G = \Mstar{} = \aplanet{} = 1$, where $\Mstar{}$ is the mass of the central star, and \aplanet{} is the planet's (initial) semi-major axis.
The planet mass, \Mplanet{}, is described in terms of the ratio of the planet mass to the stellar mass, $q \def \Mplanet{}/\Mstar{}$.
The grid extends between $r_{\text{min}} = 0.2$ and $r_{\text{max}} = 5.0$, and a full circle in azimuth. 
We used 384 equally spaced cells in the azimuthal direction, and 200 logarithmically spaced cells in the radial direction. 
This choice allows the grid cells to be nearly square and so minimises truncation errors.
The planet is initially placed at apocenter.

We initially adopted a grid identical to that of the hydrocode comparison of \texttt{de Val Borro et al.,2006 MNRAS in press}, but found that spiral density waves were spuriously reflected by the outer boundary when the planet was on an eccentric orbit.
We increased the maximum radius to that listed above and verified that our numerical experiments were not sensitive to the location of the outer boundary.

Our disk viscosity is constant over the entire disc.
We typically choose $\nu = 10^{-5}$ in the system of units outlined above (although we have performed runs with other values).
Once the viscosity is set, we may calculate the Reynolds number of the disc:
\begin{equation}
\mathcal R \equiv \frac{r^{2}\Omega}{\nu}
\label{eq:Reynoldsnumber}
\end{equation}  
In our units, $\mathcal{R} = \nu^{-1}$ at the radius of the planet's orbit (note that $\mathcal{R}$ is a function of radius).

We use a constant aspect ratio disc, with $h/r = 0.05$, where $h$ is the density scale height.
The local sound speed is set from this.
For our initial conditions, we took a flat surface density profile ($\Sigma(r) = \Sigma_0$).
Since the planet was not allowed to migrate, $\Sigma_0$ was only important numerically; we verified that it was sufficiently small to avoid spurious numerical effects.


\section{Gap opening in the circular case}
\label{sec:gapcriterion}

Planets above a certain mass are expected to open a gap in their natal disks.
In this section, we shall discuss the formation of gaps by planets on circular orbits.
Although this has been done by numerous authors, it is important that we first characterise the behaviour of \textsc{Fargo} for planets in a circular orbit, before we proceed to the eccentric case.

Previous studies have shown theoretically and numerically that a gap should be opened in a disk when the planet mass ratio exceeds
\begin{equation}
q > 40 \mathcal{R}^{-1}
\label{eq:gapopencircular}
\end{equation}
\citep{1993prpl.conf..749L,1999ApJ...514..344B}.
This condition can be derived by balancing the viscous torques, $T_{\nu} = 3 \pi \Sigma \nu r^2 \Omega$, against those from the spiral density waves the planet excites at its Lindblad resonances.

\begin{figure}
\includegraphics[angle=0,width=3.5in]{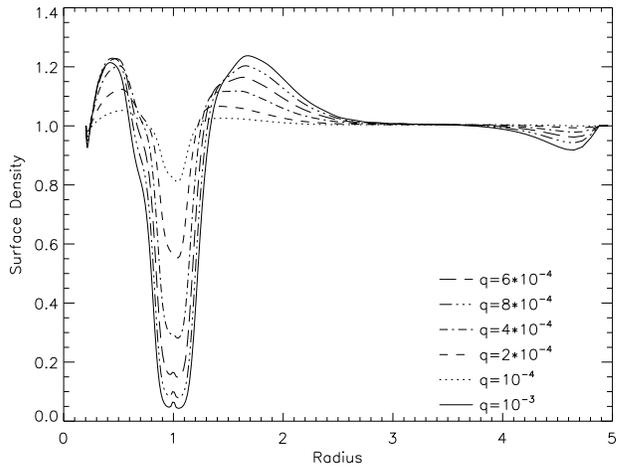}
\caption{
The azimuthally averaged surface density of the disk after 500 orbits for planet mass ratios 
$10^{-4} < q < 10^{-3}$.
All runs had the planet on a fixed circular orbits, and the disc viscosity was $\nu = 10^{-5}$}
\label{fig:viscosity}
\end{figure}

In Figure~\ref{fig:viscosity} we show the azimuthally averaged surface density after 500 orbits for planet mass ratios ranging from $q = 10^{-4}$ to $10^{-3}$ on fixed circular orbits.
These runs all have $\nu = 10^{-5}$ for their viscosity. 
We see that the depth of the gap varies smoothly with $q$, so determining whether a gap has formed is somewhat arbitrary. 
Evaluating equation~\ref{eq:gapopencircular}, we find that $q = \scinot{4}{-4}$ is the expected threshold mass ratio for gap formation for this set of runs.
Figure~\ref{fig:viscosity} shows that mass ratios larger than this have density contrasts $\SigmaGap{} / \Sigma_0 < 0.2$ at the bottom of their induced gaps.

\begin{figure}
\includegraphics[angle=0,width=3.5in]{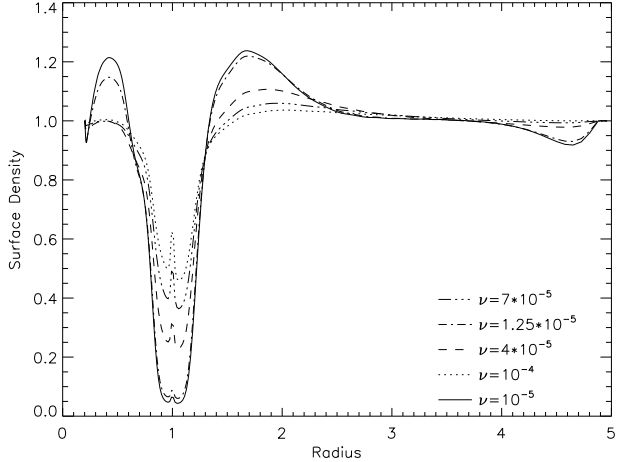}
\caption{
The azimuthally averaged surface density of the disk as the viscosity varied between $\nu = 10^{-5}$ and $\scinot{7}{-5}$.
The deeper gaps correspond to the lower viscosity disks.
All simulations are shown after 500 orbits, for a $q=10^{-3}$ planet on a circular orbit
}
\label{fig:masses}
\end{figure}

Having tested the behaviour of the code as $q$ varies, we now turn our attention to the variation with viscosity.
Figure~\ref{fig:masses} is similar to Figure~\ref{fig:viscosity}, except the planet mass ratio is fixed at $q = 10^{-3}$, and we varied the disk viscosity. 
Equation~\ref{eq:gapopencircular} predicts that for a protoplanet with $q = 10^{-3}$, a disk with Reynolds number at the planet above $\mathcal{R}=\scinot{8}{4}$ should cause a gap to open in the disk.
This corresponds to a viscosity of $\nu < \scinot{2.5}{-5}$.
The runs satisfying this criterion all show $\SigmaGap{} / \Sigma_0 < 0.2$ for their gaps.

Based on the surface density profiles shown in Figures~\ref{fig:viscosity} and~\ref{fig:masses}, we adopt the following criterion to identify formation of a gap in our numerical experiments: 
the surface density in the gap must be less than $20\%$ of the unperturbed surface density. 
We have chosen this criterion so that it is consistent with equation \ref{eq:gapopencircular}.
We adopt this criterion during our discussion of the effects of eccentric planets in subsequent sections.
However, as Figures~\ref{fig:viscosity} and~\ref{fig:masses} show, the precise definition of `gap' is arbitrary, since the variation with $q$ and $\nu$ is relatively smooth.


\section{Disk Response to an Eccentric Planet}
\label{sec:eccgapdescribe}

In this section we discuss how a disk responds to a planet on an eccentric orbit. 
We concentrate on planets with mass ratio $q = \scinot{6}{-4}$ and $10^{-3}$ in disks with viscosity $\nu = 10^{-5}$.


\subsection{Azimuthally averaged density profiles}

\begin{figure}
\includegraphics[angle=0,width=3.5in]{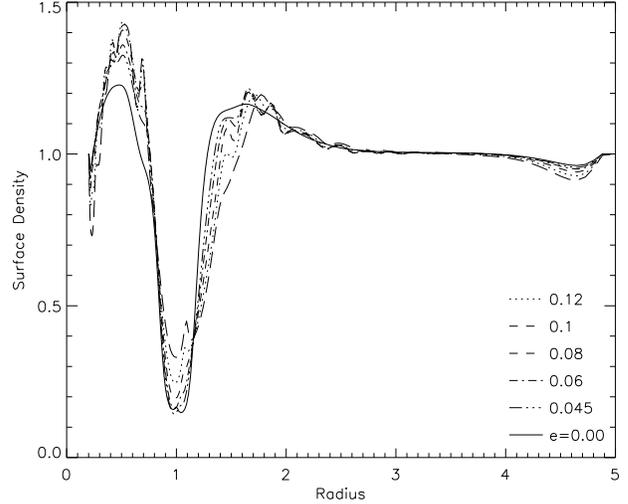}
\caption{
The azimuthally averaged surface density of the disk after 500 orbits for planet eccentricities between  $0 < \eplanet{} < 0.12$. 
The planets had $q= 6 \times 10^{-4}$ and the disc viscosity was $\nu = 10^{-5}$}
\label{fig:mass6}
\end{figure}

In Figure~\ref{fig:mass6} we show how the surface density profiles vary with planetary eccentricity, \eplanet{} for a $q = \scinot{6}{-4}$ planet.
We see that the shape of the gap is similar to the circular case for planet eccentricity below $e < 0.06$.
The planet's Hill radius is given by
\begin{equation}
r_{\text{Hill}} = \parenfrac{q}{3}^{\frac{1}{3}} \aplanet{}
\label{eq:HillSphereDefine}
\end{equation}
which measures the distance over which the planet's gravity dominates (the Hill sphere is the size of the Roche lobe in the low $q$ limit).
In our units, we note that $r_{\text{Hill}} = 0.058$ for these runs.
Above this eccentricity, the gap as seen in the azimuthally averaged density profiles, is shallower and wider. 
By the 20\% criterion discussed in the previous section, the planet no longer induces a gap for $\eplanet{} > 0.1$.

\begin{figure}
\includegraphics[angle=0,width=3.5in]{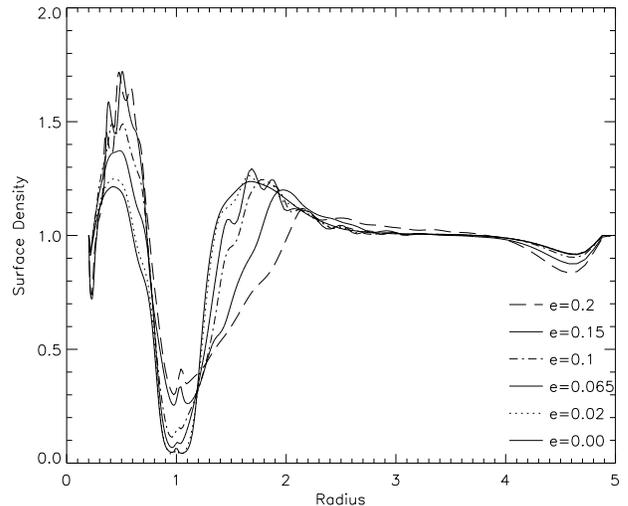}
\caption{
Similar to Figure~\ref{fig:mass6}, but the planet is mass ratio $q=10^{-3}$ and the eccentricity range is $0< e_p < 0.2$}
\label{fig:jup500}
\end{figure}

Figure~\ref{fig:jup500} is similar to Figure~\ref{fig:mass6} but for a $q=10^{-3}$ planet.
At eccentricities of $\eplanet{} < 0.065$, the shape of the gap is nearly identical to that of the circular case.
Above this threshold, the gap becomes shallower and wider, and increasingly deviates from the circular case. 
We note that the planet's Hill radius is $0.069 \aplanet{}$.
This suggests that the planet's eccentricity has little effect on the density profile unless the planet's orbit takes it outside the sum of its semi-major axis and Hill radius.
As spiral density waves are driven from resonances located at radii all the way up to the planet's Hill radius, this limiting eccentricity is not unexpected.

Figures~\ref{fig:mass6} and~\ref{fig:jup500} show that as planet eccentricity increases, the azimuthally averaged gap density profile becomes wider and shallower.
It becomes harder for the protoplanet to open up a deep gap in the disk. 
Because the density in the gap is lower at higher planet eccentricity, a more massive planet is required to open a gap in the disk.
We shall discuss possible explanations for this behaviour after we describe the disk morphology seen in these simulations.

\subsection{Eccentric gaps}

\begin{figure}
\subfigure[ ]{
\includegraphics[angle=0,width=3.5in]{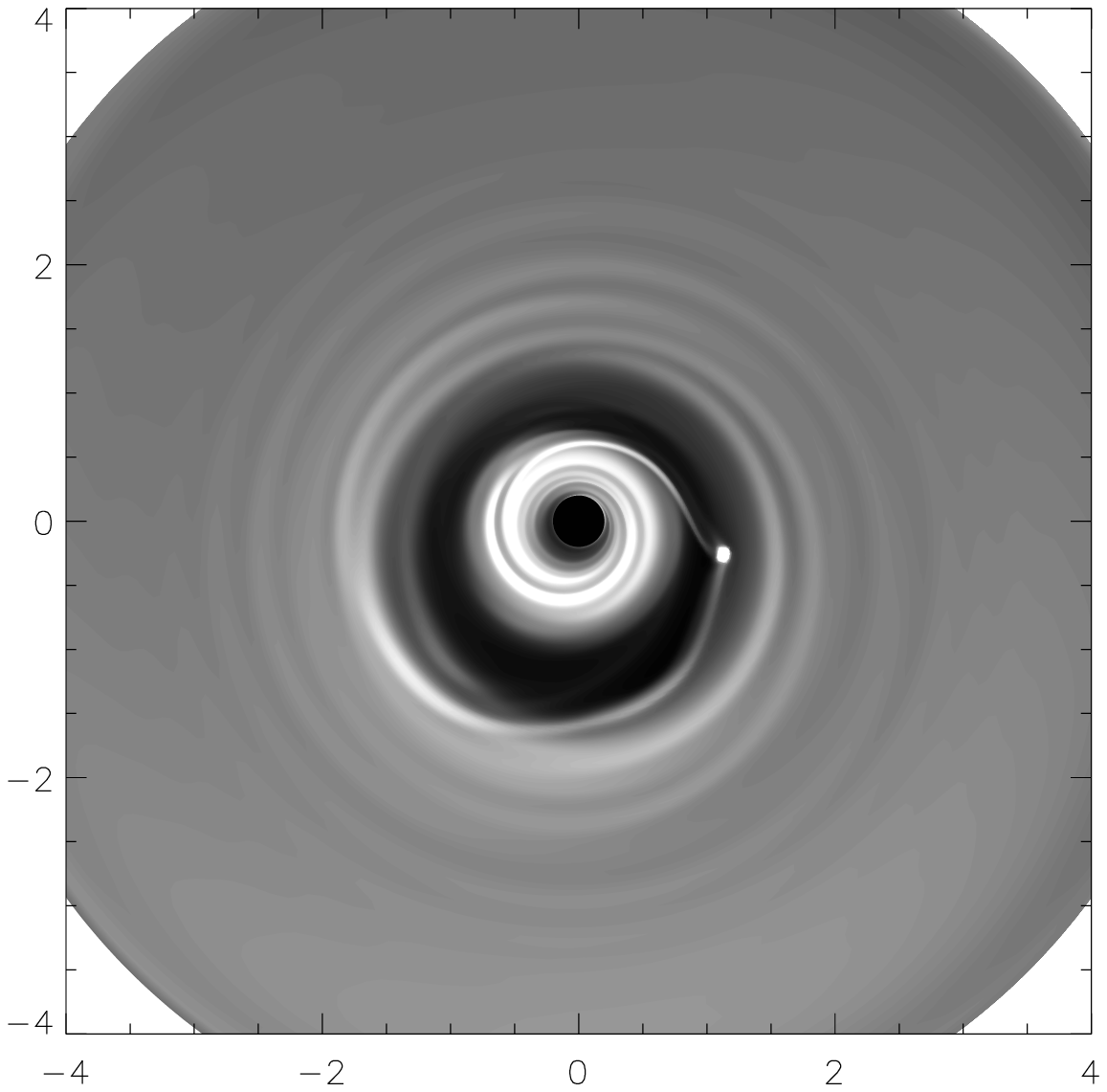}
}
\subfigure[ ]{
\includegraphics[angle=0,width=3.5in]{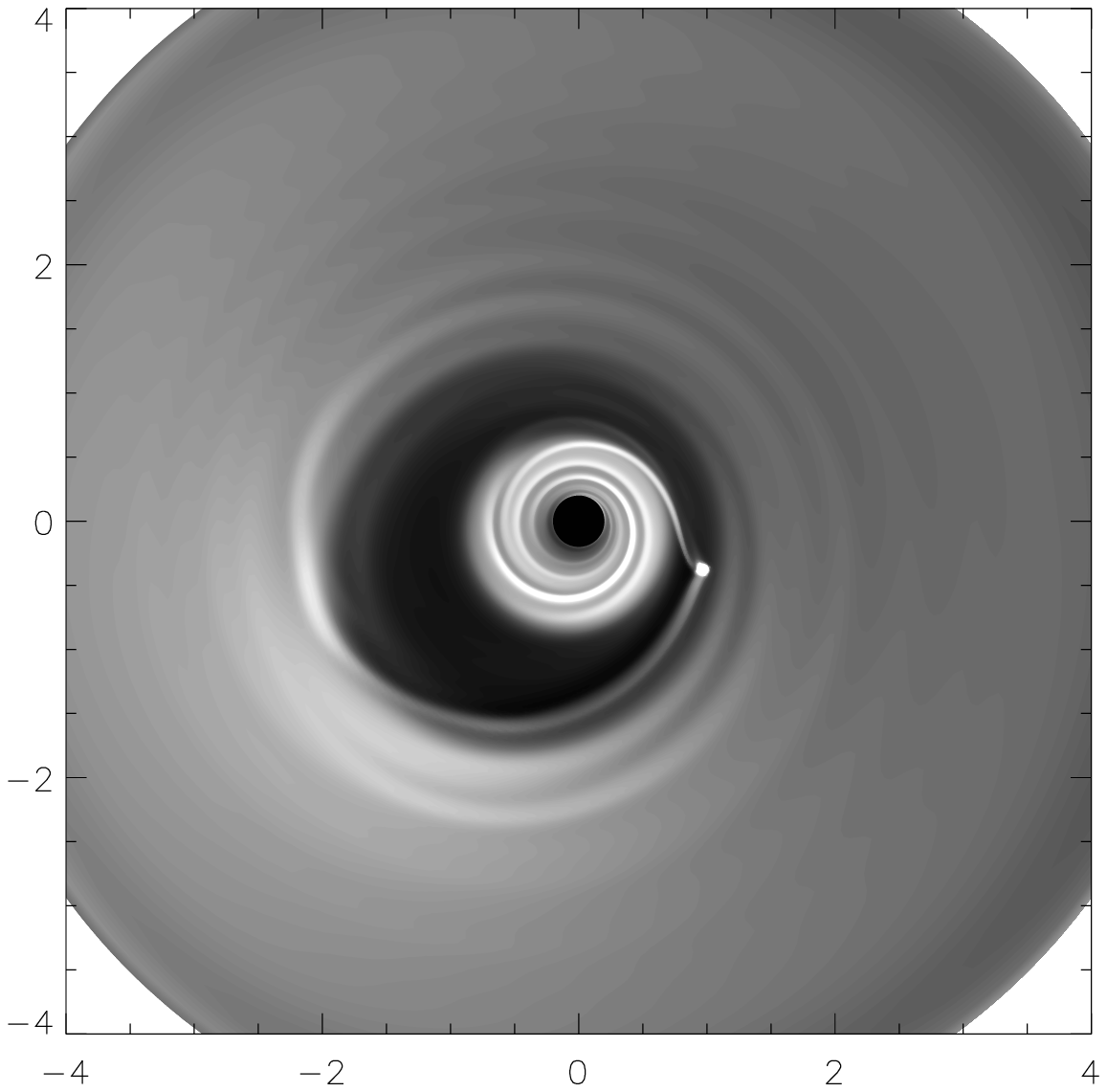}
}
\caption{Disc surface densities after a) 250 and b) 500 orbits for a $q=10^{-3}$ planet on a orbit with eccentricy fixed at $\eplanet{} =0.2$.
The gas viscosity was $\nu = 10^{-5}$}
\label{fig:contour_jup4}
\end{figure}

In Figure~\ref{fig:contour_jup4}, we show the 2D surface density of the disc after 250 and 500 orbits for planet eccentricity $\eplanet{} = 0.2$ and mass ratio $q=10^{-3}$.
Examining the surface density as a function of time, we see the planet crossing between the inner and outer gap edges, producing one armed spirals as it encounters each edge.    
By $t=500$ orbits, the planet has formed an eccentric gap.
The eccentricity of the gap increases to a maximum of about twice that of the planet at this time.
The gap also slowly precesses.
At $t=500$ orbits the gap's apocentre is about $180^{\circ}$ from that of the planet.

Because the gap precesses with respect to the planet's orbit, the planet
approaches the disk edge at different longitudes.
Close approaches allow the planet to clear out additional material from the disk edge.
On longer timescales, $t > 1000 $ orbits, the gap slowly widens and eventually circularizes.  
However during the long circularization period accretion onto the planet could exceed that of a similar mass planet on a circular orbit.  
In the case of a planet on a circular orbit, the formation of a gap can significantly reduce the accretion rate onto the planet \citep{2003ApJ...586..540D}.
Recently, \citet{2006A&A...447..369K} looked at the excitation of disc eccentricity by massive planets (up to $q=\scinot{5}{-3}$), and suggested that the induced eccentricity could aid accretion by the planet.
Although we did not permit accretion in our numerical experiments, we would expect behaviour of this nature.

Figure~\ref{fig:jup500} showed the azimuthally averaged density profile was shallower for gaps opened by eccentric planets.
The smoothness or shallow gradient of the azimuthally averaged profile in the gap is due to the eccentricity of the gap edge, rather than a smooth change in the disk density with radius.
The gap itself has sharp edges.  
The apparent shallow profile in the azimuthally averaged profiles is caused by averaging the density profile at different longitudes.

\begin{figure}
\includegraphics[angle=0,width=3.5in]{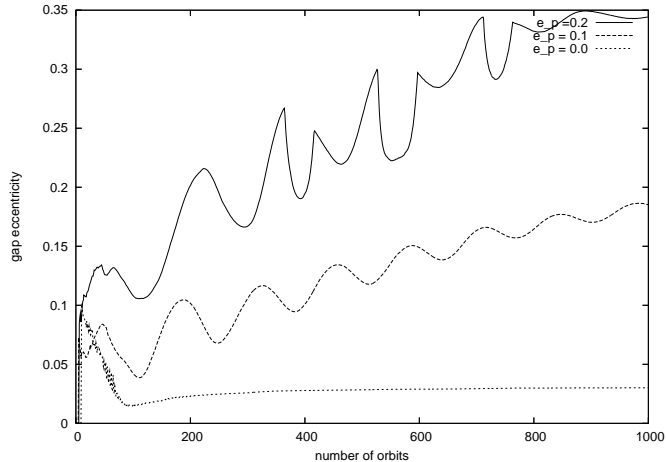}
\caption{The eccentricity of the gap opened in the disk as a function of time for a $q=10^{-3}$ planet, for $\eplanet{}=0$, $0.1$ \&~$0.2$, and a disk viscosity of $\nu=10^{-5}$.
We computed the eccentricity by locating all the cells satisfying $\Sigma / \Sigma_{0} < 0.2$, treating each as a free particle, and taking the mean of the resulting eccentricities
}
\label{fig:GapEccenEvolve}
\end{figure}

Figure~\ref{fig:GapEccenEvolve} plots the eccentricity evolution of the gap for three values of \eplanet{}.
To determine the eccentricity of the gap, we selected cells based on the $\Sigma / \Sigma_{0} < 0.2$ criterion introduced in section~\ref{sec:gapcriterion}.
We then computed the cell eccentricity by treating each as a free particle, and took the mean of these.
Because of the rather arbitrary gap criterion used, the time evolution shows some large oscillations.\footnote{We could produce a weighted mean, but the appropriate weighting is not obvious, so we prefer to show the simple average}
The circular ($\eplanet{}=0$) case shows a brief initial spike of eccentricity in the gap, before decaying to a low value.
For the two runs with an eccentric planet, we see a steady growth in gap eccentricity, approaching $2 \eplanet{}$ after 1000 orbits.

\section{Gap opening as a function of planet mass and eccentricity}
\label{sec:eccgapcriterion}

\begin{figure}
\includegraphics[angle=0,width=3.0in]{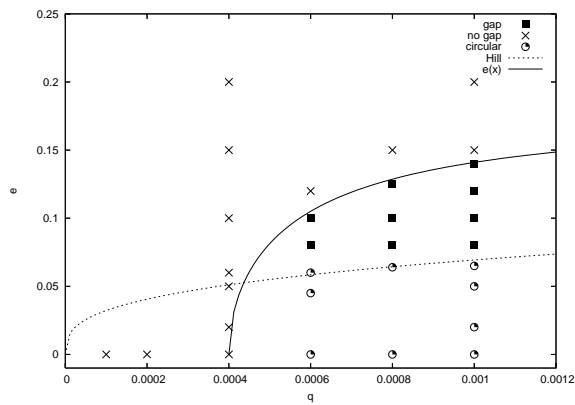}
\caption{
Gap formation as a function of planet mass ratio and eccentricity. 
For these runs the disk viscosity was $\nu = 10^{-5}$ so a gap is expected for $q > 4 \times 10^{-4}$.
The shaded square points refer to runs in which a gap was opened in the disk.
The empty square points represent runs that closely followed their respective circular cases when a gap was formed.
The $\times$ indicates a run lacking a gap.
The dotted line shows the planet's Hill radius divided by the planet's semi-major axis.
Simulations below this line closely resembled their circular orbit counterparts.
At large planet eccentricities, the planet could not open a gap. 
We have empirically fit a function, shown as a dashed line to mark
the division between gap opening simulations and those lacking a gap.
This function is in the form (equation~\ref{eq:fitequation}) expected if co-rotation resonances reduce the total torque on the disk due to density waves driven by the planet
}
\label{fig:etrace3}
\end{figure}

We performed a series of runs, all with $\nu = 10^{-5}$, and with different planet masses and eccentricities.
For each run we measured the depth of the density deficit near the planet and labeled the simulation as ``gap opening'' if the density near the planet was lower than 20\% of the unperturbed value. 
Figure~\ref{fig:etrace3} shows the result of these runs.
We distinguish between three behaviours:
no gap opened (by the 20\% criterion introduced above);
a gap formed which is significantly different to the circular case;
and a gap formed which is essentially identical to the circular case.

From Figure~\ref{fig:etrace3}, we see that simulations similar to the circular case are found at the bottom of the plot.
The division between density profiles resembling the circular case and those differing is not a strong function of eccentricity. 
The border between the two regions is well described by
\begin{equation}
e_{\textrm{threshold}} \approx  \parenfrac{q}{3}^{\frac{1}{3}}
\label{eq:threshold_eccen}
\end{equation}
This is essentially equation~\ref{eq:HillSphereDefine}, and compares the variation in peri- and apastron to the planet's Hill sphere. 
This suggests that the planet's eccentricity has little effect on the density profile unless the planet's orbit takes it outside the sum of its semi-major axis and Hill radius.
Spiral density waves are driven by resonances up to the edge of the planet's Hill sphere (this assumption is used to derive equation~\ref{eq:gapopencircular}), so it is not unsurprising that once the planet strays over these resonances (effectively destroying them) the behaviour changes.

Figure~\ref{fig:etrace3} shows that at high planet eccentricity a gap is not opened in the disk even when equation~\ref{eq:gapopencircular} would predict one.
We now discuss possible reasons for this.
We expect a gap is opened when torque due to damping of density waves exceeds that due to viscous flow. 
We first consider the accretion rate through a viscous and eccentric disk.
\citet{1992MNRAS.255...92S,1993MNRAS.260..463S} showed that the torque due to a viscous accretion varies with eccentricity according to $T_\nu \propto (1-e^2)$ (see their equation 2.16).
However this reduces the viscous torque rather than increasing it, and would make it easier for an eccentric planet to open a gap rather than harder as we see from Figure~\ref{fig:etrace3}.

The solution lies in the other reseonances present in the system.
Because of perturbations from the planet, spiral density waves are excited at both Lindblad and co-rotation resonances.
Lindblad resonances are expected to increase the eccentricity of the planet, and hence would decrease the eccentricity of the disk \citep[e.g.][]{2004ApJ...606L..77S}.
Co-rotation resonances, however, when in the presence of a density gradient, damp the eccentricity of the planet and hence are expected to increase the eccentricity of the disk.
First order terms in the potential, proportional to the planet eccentricity, can drive co-rotation resonances into the disk at the same location as the traditional Lindblad resonances.
Notably the corotation resonances are also expected to have torque with the opposite sign as the Lindblad resonances.
Waves driven at these resonances and damped in the disk do not push the disk away from the planet, but rather pull the disk toward the planet.
Consequently the corotation resonances reduce the torque on the disk due to spiral density waves driven by the planet.
The torque from these waves $T_{CR} \propto \deriv{\Sigma}{r} q^2 \eplanet{2}$  \cite[from equation~14 of][]{1980ApJ...241..425G}
We therefore expect a modified gap opening criterion of the following form
\begin{equation}
q \paren{1 - \chi \eplanet{2}} = 40 \mathcal{R}^{-1}
\label{eq:fitequation}
\end{equation}
where $\chi$ depends on the density gradient and the ratio between the sum of torques from the corotation resonances  and the sum of torques from the Lindblad  resonances.
Unfortunately $\chi$ is dependent on the density profile and is not easy to calculate.
The parameter $\chi$ may be large since a sharp gradient in the density could cause a large torque even if the gap is not deep.
Since the shape of the gap is certain to be a function of viscosity (cf Figure~\ref{fig:masses}), we expect $\chi$ to vary with viscosity too.
However, we did not vary viscosity in this set of runs.

We have fit a function in the form equation~\ref{eq:fitequation} consistent with the gap/no-gap line seen in Figure~\ref{fig:etrace3}.
The gap/no-gap curve is consistent with $\chi \approx 912$.
This  suggests that the corotation resonances are primarily important in regions of strong density gradients.   
Only a planet massive enough to cause a moderate density gradient in the disk would be able to increase the disk eccentricity.


\section{Discussion}
\label{sec:discuss}

We have seen that an eccentric planet can open an eccentric gap in a disc.
The eccentricity should be an observable quantity.
Although we see eccentricity in images of discs - the eccentric ring around Fomalhaut \citep{2005Natur.435.1067K} for example - we would expect eccentric holes of the nature described here to only be a few AU from their host stars.
Such holes would be too small to be imaged by current observatories.
However, a strongly eccentric gap should leave a mark on the spectral energy distribution (SED) of a disk.
The surface temperature of a disk illuminated by the star is set by the radial distance $T\propto r^{-1/2}$.   
Consequently the eccentric edge of a disk would correspond to a range of dust temperatures in the edge.
The spectral energy distribution of CoKuTau/4's disk edge was consistent with a single temperature optically thick wall \citep{2005ApJ...621..461D}.
However an eccentric hole would produce a spectral energy distribution consistent with a range of dust edge temperatures.  
Such a spectrum could either be interpreted as due to a smooth radial density gradient or an eccentric hole.
If the disk were optically thick then the second possibility would be more likely than the first.
We have found that the disk edge can be more highly eccentric than the planet, consequently the variation in edge temperature around an eccentric disk edge may be detectable even though the dust temperature is only proportional to the inverse of the square root of the radius.

How might such a system form?
In our runs, the planet was on a fixed orbit.
\citet{2001A&A...366..263P} note that the planet and disc should exchange eccentricity, with the exchange being most efficient when the planet and disc are of comparable mass.
A massive disk would also be expected to damp the planet's eccentricity \citep{1993ApJ...419..166A}.
We are therefore lead to two possible situations in which an eccentric gap might arise.
The first is the case of an eccentric planet in an old, dissipating disk.
In this case, we expect the planet to be more massive than the disk, and hence little affected by it.
How the planet achieved its high eccentricity is then an open question.
The second scenario requires a second, inner planet.
If locked in resonance with the outer planet, it could `resupply' eccentricity to the outer planet, even in the face of damping by the outer disk.
We might even expect the outer planet to have been pushed into the resonance by conventional planet-disk migration.
This is somewhat analagous to the case of GJ~876, as discussed by \citet{2005A&A...437..727K}.
However, we would expect the inner planet to be more massive, and hence less affected by the resonant interaction.
This is not the case in GJ~876, where the inner planet is believed to be lower mass, and on a substantially more eccentric orbit.
\citet{2005A&A...437..727K} required rapid dissipation of their gas disk, to prevent the inner planet becoming too eccentric.


\section{Summary}
\label{sec:summary}

In this paper, we have examined the effect of an eccentric planet on a gas disk.
From our numerical experiments, we propose that the conventional gap-opening criterion should be modified to become
\begin{equation}
q \paren{1 - 912 \eplanet{2}} = 40 \mathcal{R}^{-1}
\label{eq:newgapcriterion}
\end{equation}
The form of equation~\ref{eq:newgapcriterion} is justified by consideration of corotation torques, which are expected to act to oppose the Lindblad torques.
However, the extra coefficient had to be determined numerically, due to the sensitivity of the corotation torques to the exact gap profile.
We expect its value to depend on the assumed disc viscosity, but we have not parameterised this variation.

We have found that the gap eccentricity can substantially exceed the planet's eccentricity.
In the early stages of evolution, as the planet crosses between the inner and outer discs, it induces one armed spirals in each.
Coupled with the precession of the gap, we expect this to lead to accretion by the planet, despite the presence of a gap.

If planet's eccentricity is sufficiently low, then the gap formed is almost identical to the circular case.
We have found the threshold eccentricity to be well fit by
\begin{equation}
\eplanet{} < \parenfrac{q}{3}^{\frac{1}{3}}
\label{eq:circularthreshold}
\end{equation}
which compares the peri- and apocentre of the planet to its Hill sphere.
The form of equation~\ref{eq:circularthreshold} is compatible with the standard theory of gap formation.
This theory assumes that resonances up to the edge of the Hill sphere contribute to opening the gap (within the Hill sphere, the resonances do not exist, since there material orbits the planet, not the star).
Once \eplanet{} is high enough to let the planet move outside its notional Hill sphere, these resonances will be destroyed, and we would expect the morphology of the gap to change.

We would expect an eccentric gap to be detectable in a disk SED.
However, without corroborating data, it might be difficult to disentangle eccentricity from an azimuthally symmetric radial density gradient.

Future work is needed, allowing the planet to feel the disk.
This will allow us to determine the maximum disk mass which does not rapidly circularize the planet.
We have also suggested that a high planetary eccentricity might be maintained by a second, massive, inner planet, even in the presence of strong damping by a higher mass disk.
Further calculations are needed to examine this possibility.


\bibliography{general}

\begin{thebibliography}{}

\bibitem[\protect\astroncite{{Artymowicz}}{1993}]{1993ApJ...419..166A}
{Artymowicz}, P.: 1993,
\newblock {\em \apj} {\bf 419}, 166

\bibitem[\protect\astroncite{{Bergin} et~al.}{2004}]{2004ApJ...614L.133B}
{Bergin}, E., {Calvet}, N., {Sitko}, M.~L., {Abgrall}, H., {D'Alessio}, P.,
  {Herczeg}, G.~J., {Roueff}, E., {Qi}, C., {Lynch}, D.~K., {Russell}, R.~W.,
  {Brafford}, S.~M., and {Perry}, R.~B.: 2004,
\newblock {\em \apjl} {\bf 614}, L133

\bibitem[\protect\astroncite{{Bryden} et~al.}{1999}]{1999ApJ...514..344B}
{Bryden}, G., {Chen}, X., {Lin}, D.~N.~C., {Nelson}, R.~P., and {Papaloizou},
  J.~C.~B.: 1999,
\newblock {\em \apj} {\bf 514}, 344

\bibitem[\protect\astroncite{{Calvet} et~al.}{2005}]{2005ApJ...630L.185C}
{Calvet}, N., {D'Alessio}, P., {Watson}, D.~M., {Franco-Hern{\'a}ndez}, R.,
  {Furlan}, E., {Green}, J., {Sutter}, P.~M., {Forrest}, W.~J., {Hartmann}, L.,
  {Uchida}, K.~I., {Keller}, L.~D., {Sargent}, B., {Najita}, J., {Herter},
  T.~L., {Barry}, D.~J., and {Hall}, P.: 2005,
\newblock {\em \apjl} {\bf 630}, L185

\bibitem[\protect\astroncite{{Cresswell} and
  {Nelson}}{2006}]{2006A&A...450..833C}
{Cresswell}, P. and {Nelson}, R.~P.: 2006,
\newblock {\em \aap} {\bf 450}, 833

\bibitem[\protect\astroncite{{D'Alessio} et~al.}{2005}]{2005ApJ...621..461D}
{D'Alessio}, P., {Hartmann}, L., {Calvet}, N., {Franco-Hern{\'a}ndez}, R.,
  {Forrest}, W.~J., {Sargent}, B., {Furlan}, E., {Uchida}, K., {Green}, J.~D.,
  {Watson}, D.~M., {Chen}, C.~H., {Kemper}, F., {Sloan}, G.~C., and {Najita},
  J.: 2005,
\newblock {\em \apj} {\bf 621}, 461

\bibitem[\protect\astroncite{{D'Angelo} et~al.}{2003}]{2003ApJ...586..540D}
{D'Angelo}, G., {Kley}, W., and {Henning}, T.: 2003,
\newblock {\em \apj} {\bf 586}, 540

\bibitem[\protect\astroncite{{Goldreich} and
  {Sari}}{2003}]{2003ApJ...585.1024G}
{Goldreich}, P. and {Sari}, R.: 2003,
\newblock {\em \apj} {\bf 585}, 1024

\bibitem[\protect\astroncite{{Goldreich} and
  {Tremaine}}{1980}]{1980ApJ...241..425G}
{Goldreich}, P. and {Tremaine}, S.: 1980,
\newblock {\em \apj} {\bf 241}, 425

\bibitem[\protect\astroncite{{Jensen} and
  {Mathieu}}{1997}]{1997AJ....114..301J}
{Jensen}, E.~L.~N. and {Mathieu}, R.~D.: 1997,
\newblock {\em \aj} {\bf 114}, 301

\bibitem[\protect\astroncite{{Kalas} et~al.}{2005}]{2005Natur.435.1067K}
{Kalas}, P., {Graham}, J.~R., and {Clampin}, M.: 2005,
\newblock {\em \nat} {\bf 435}, 1067

\bibitem[\protect\astroncite{{Kley} and {Dirksen}}{2006}]{2006A&A...447..369K}
{Kley}, W. and {Dirksen}, G.: 2006,
\newblock {\em \aap} {\bf 447}, 369

\bibitem[\protect\astroncite{{Kley} et~al.}{2005}]{2005A&A...437..727K}
{Kley}, W., {Lee}, M.~H., {Murray}, N., and {Peale}, S.~J.: 2005,
\newblock {\em \aap} {\bf 437}, 727

\bibitem[\protect\astroncite{{Kley} et~al.}{2004}]{2004A&A...414..735K}
{Kley}, W., {Peitz}, J., and {Bryden}, G.: 2004,
\newblock {\em \aap} {\bf 414}, 735

\bibitem[\protect\astroncite{{Lin} and
  {Papaloizou}}{1993}]{1993prpl.conf..749L}
{Lin}, D.~N.~C. and {Papaloizou}, J.~C.~B.: 1993,
\newblock in E.~H. {Levy} and J.~I. {Lunine} (eds.), {\em Protostars and
  Planets III}, pp 749--835

\bibitem[\protect\astroncite{{Marsh} and {Mahoney}}{1992}]{1992ApJ...395L.115M}
{Marsh}, K.~A. and {Mahoney}, M.~J.: 1992,
\newblock {\em \apjl} {\bf 395}, L115

\bibitem[\protect\astroncite{{Masset}}{2000a}]{2000A&AS..141..165M}
{Masset}, F.: 2000a,
\newblock {\em \aaps} {\bf 141}, 165

\bibitem[\protect\astroncite{{Masset} and
  {Snellgrove}}{2001}]{2001MNRAS.320L..55M}
{Masset}, F. and {Snellgrove}, M.: 2001,
\newblock {\em \mnras} {\bf 320}, L55

\bibitem[\protect\astroncite{{Masset}}{2000b}]{2000ASPC..219...75M}
{Masset}, F.~S.: 2000b,
\newblock in {\em ASP Conf. Ser. 219: Disks, Planetesimals, and Planets}, p.~75

\bibitem[\protect\astroncite{{Papaloizou}}{2003}]{2003CeMDA..87...53P}
{Papaloizou}, J.~C.~B.: 2003,
\newblock {\em Celestial Mechanics and Dynamical Astronomy} {\bf 87}, 53

\bibitem[\protect\astroncite{{Papaloizou} et~al.}{2001}]{2001A&A...366..263P}
{Papaloizou}, J.~C.~B., {Nelson}, R.~P., and {Masset}, F.: 2001,
\newblock {\em \aap} {\bf 366}, 263

\bibitem[\protect\astroncite{{Papaloizou} and {Terquem}}{2006}]{papa06}
{Papaloizou}, J.~C.~B. and {Terquem}, C.: 2006,
\newblock {\em Reports on Progress in Phys.} {\bf 69}, 119

\bibitem[\protect\astroncite{{Rice} et~al.}{2003}]{2003MNRAS.342...79R}
{Rice}, W.~K.~M., {Wood}, K., {Armitage}, P.~J., {Whitney}, B.~A., and
  {Bjorkman}, J.~E.: 2003,
\newblock {\em \mnras} {\bf 342}, 79

\bibitem[\protect\astroncite{{Sari} and
  {Goldreich}}{2004}]{2004ApJ...606L..77S}
{Sari}, R. and {Goldreich}, P.: 2004,
\newblock {\em \apjl} {\bf 606}, L77

\bibitem[\protect\astroncite{{Syer} and {Clarke}}{1992}]{1992MNRAS.255...92S}
{Syer}, D. and {Clarke}, C.~J.: 1992,
\newblock {\em \mnras} {\bf 255}, 92

\bibitem[\protect\astroncite{{Syer} and {Clarke}}{1993}]{1993MNRAS.260..463S}
{Syer}, D. and {Clarke}, C.~J.: 1993,
\newblock {\em \mnras} {\bf 260}, 463

\bibitem[\protect\astroncite{{Thommes} and
  {Lissauer}}{2003}]{2003ApJ...597..566T}
{Thommes}, E.~W. and {Lissauer}, J.~J.: 2003,
\newblock {\em \apj} {\bf 597}, 566

\end{thebibliography}
\bibliographystyle{astron}


\section*{Acknowledgements}

Support for this work was in part provided by National Science Foundation grants AST-0406823 \& PHY-0552695, and the National Aeronautics and Space Administration under Grant No.~NNG04GM12G issued through the Origins of Solar Systems Program, and HST-AR-10972 from the Space Telescope Science Institute.
We thank Dan Watson for reading an early version of this manuscript, and providing several useful comments.

\bsp

\label{lastpage}

\end{document}